\documentclass[12pt,a4paper]{article}

\usepackage[text={16.5cm,23cm},centering]{geometry}
\usepackage{amsmath}
\usepackage{comment}
\usepackage{graphicx}
\usepackage{graphbox}
\graphicspath{{figure/}}
\usepackage[%
  format=plain,
  margin=1cm,
  font=small
]{caption}
\usepackage{xspace}
\usepackage[utf8]{inputenc}
\usepackage{tgtermes,tgheros,tgcursor}
\usepackage[varg]{newtxmath}
\usepackage{hyperref}
\usepackage{subcaption}
\hypersetup{colorlinks=true,linkcolor=blue,citecolor=blue}
\usepackage[nosort]{cite}

\numberwithin{equation}{section}

\newcommand{\Zb}{\mathbb{Z}}
\newcommand{\bbb}{quarter 16-cell\xspace}
\newcommand{\cubebb}{cubic cone\xspace}
\newcommand{\deltamod}[1]{\delta^{\mathrm{mod}\ 2}_{#1,0}}
\newcommand{\diskv}{$D^3$ expectation value\xspace}
\newcommand{\diskvs}{$D^3$ expectation values\xspace}
\newcommand{\Dtb}{$\widetilde{\mathrm{D}}$\xspace}
\newcommand{\Dt}{\widetilde{\mathrm{D}}}
\newcommand{\Z}[1]{Z_{\text{#1}}}

\newcommand{\bsD}{s_{\text{D}}}
\newcommand{\blD}{l_{\text{D}}}
\newcommand{\bWD}{W_{\text{D}}}
\newcommand{\bsDt}{s_{\widetilde{\mathrm{D}}}}
\newcommand{\blDt}{l_{\widetilde{\mathrm{D}}}}
\newcommand{\bWDt}{W_{\widetilde{\mathrm{D}}}}
\newcommand{\pD}{p_{\text{D}}}
\newcommand{\pDt}{p_{\widetilde{\mathrm{D}}}}
\newcommand{\Q}[2]{Q(\mbox{#1};\mbox{#2})}
\newcommand{\g}[1]{g_{#1}}
\newcommand{\gd}{g_{\mathrm{D}}}
\newcommand{\gn}{g_{\mathrm{N}}}
\newcommand{\gdt}{g_{\widetilde{\mathrm{D}}}}

\begin{document}
\begin{center}
\begin{flushright}
  OU-HET 1180
\end{flushright}
\vspace{8ex}
{\Large \bfseries \boldmath Non-invertible symmetries and boundaries in four dimensions}\\
\vspace{4ex}
{\Large Masataka Koide, Yuta Nagoya, and Satoshi Yamaguchi}\\
\vspace{2ex}
{\itshape Department of Physics, Graduate School of Science, 
\\
Osaka University, Toyonaka, Osaka 560-0043, Japan}\\
\vspace{1ex}
Email: \texttt{\small mkoide@het.phys.sci.osaka-u.ac.jp, y\_nagoya@het.phys.sci.osaka-u.ac.jp, yamaguch@het.phys.sci.osaka-u.ac.jp}\\
\begin{abstract}
We study quantum field theories with boundaries by utilizing non-invertible symmetries.
We consider three kinds of boundary conditions of the four dimensional $\Zb_2$ lattice gauge theory at the critical point as examples.
The weights of the elements on the boundary are determined so that these boundary conditions are related by the Kramers-Wannier-Wegner (KWW) duality.
In other words, it is required that the KWW duality defects ending on the boundary are topological.
Moreover, we obtain the ratios of the hemisphere partition functions with these boundary conditions; this result constrains the boundary renormalization group flows.
\end{abstract}
\end{center}

\vspace{4ex}
\section{Introduction}

Recently, topological defects in quantum field theories have been attracting attention, and actively studied as a generalization of symmetries \cite{Gaiotto:2014kfa}.
One class of such generalizations is non-invertible symmetry.
The non-invertible symmetries do not have any group structure, but they have structures of (higher) fusion categories\cite{Douglas:2018,Johnson-Freyd:2020}.
The study of such non-invertible symmetries has been active in two dimensions\cite{Verlinde:1988sn,Moore:1988qv,Moore:1989yh,Petkova:2000ip,Fuchs:2002cm,Frohlich:2004ef,Frohlich:2006ch,Feiguin:2006ydp,Aasen:2016dop,Bhardwaj:2017xup,Chang:2018iay,Lin:2019hks,Thorngren:2019iar,Gaiotto:2020iye,Komargodski:2020mxz,Aasen:2020jwb,Inamura:2021wuo,Thorngren:2021yso,Huang:2021zvu,Huang:2021nvb,Inamura:2022lun,Lin:2022dhv}, as well as in higher dimensions \cite{Rudelius:2020orz,Nguyen:2021yld,Heidenreich:2021xpr,Koide:2021zxj,Choi:2021kmx,Kaidi:2021xfk,Wang:2021vki,Roumpedakis:2022aik,Bhardwaj:2022yxj,Hayashi:2022fkw,Choi:2022zal,Choi:2022jqy,Cordova:2022ieu,Bhardwaj:2022lsg,Bartsch:2022mpm,Apruzzi:2022rei,GarciaEtxebarria:2022vzq,Heckman:2022muc,Kaidi:2022cpf,Freed:2022qnc,Antinucci:2022vyk,Chen:2022cyw,Cordova:2022fhg,Yokokura:2022alv,Bhardwaj:2022kot,Bhardwaj:2022maz,Bartsch:2022ytj,Heckman:2022xgu,Das:2022fho,Apte:2022xtu,Delcamp:2023kew,Kaidi:2023maf,Putrov:2023jqi,Carta:2023bqn}.  For more references on the recent developments on the generalized symmetry, see \cite{Cordova:2022ruw} and references therein.

In two-dimensional conformal field theories (CFTs), the conformal anomaly is monotonically decreasing along the renormalization group flow;
this is called the ``c-theorem'' \cite{Zamolodchikov:1986gt}.
This theorem is useful for understanding the renormalization group flow.
A similar theorem exists in four dimensions and is called the ``a-theorem'' \cite{Cardy:1988cwa,Osborn:1989td,Komargodski:2011vj,Casini:2022bsu}.
A similar statement is conjectured for conformal field theories with boundaries \cite{Affleck:1991tk,Nozaki:2012qd,Gaiotto:2014gha,Kobayashi:2018lil}; the hemisphere partition function with a given conformal boundary condition is monotonically decreasing along boundary renormalization group flow in two and three dimensions and monotonically increasing in four dimensions.
This statement is proved in two dimensions \cite{Friedan:2003yc,Casini:2016fgb}, in three dimensions \cite{Casini:2018nym}, and in four dimensions\cite{Casini:2023kyj}.
In this paper, we call the hemisphere partition function the ``g-function.''

Ordinary symmetries give relations between g-functions.
Actually, if two boundary conditions are related by an ordinary symmetry, their g-functions are identical to each other.
For example, let us consider spin-up and spin-down fixed boundary conditions, denoted by $+,-$, respectively, in the two-dimensional Ising CFT.
Their g-functions $g_{\pm}$ satisfy $g_+=g_-$ since they are related by the spin-flip symmetry.
In terms of the topological defect, this equality comes from the fact that the boundary conditions $\pm$ are related by the fusion of the spin-flip symmetry defect, whose quantum dimension is $1$. 

\begin{figure}[htbp]
  \centering
  \includegraphics[width=10cm]{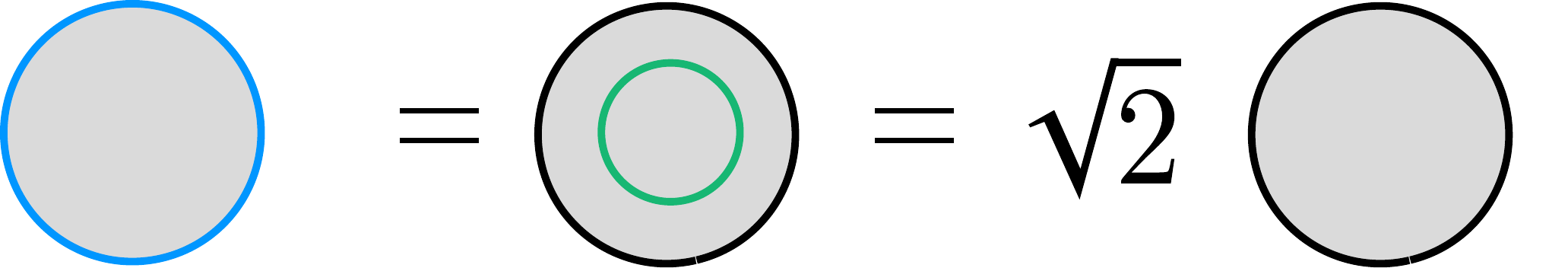}
  \caption{Derivation of the relation between the g-functions. 
The gray disks are hemispheres on which the Ising CFT lives.
The black boundary represents the fixed boundary condition $+$ and the blue boundary represents the free boundary condition $0$.
The green circle is the KW duality defect.
The KW duality defect can act on the boundary with the $+$ boundary condition, and change the boundary condition to the $0$ boundary condition (the left-hand side).
On the other hand, a circular KW duality defect that does not contain operators inside can be replaced by its quantum dimension $\sqrt{2}$ (the right-hand side).}
  \label{fig:g-function1}
\end{figure}

Non-invertible symmetries also give relations between g-functions in the same way.
For example, let us consider the two-dimensional Ising CFT again.
In addition to $\pm$ boundary conditions, we also have the free boundary condition $0$, whose g-function is denoted by $g_0$.
Now consider the Ising CFT on a two-dimensional hemisphere with $+$ boundary condition, and place on that hemisphere the non-invertible topological defect associated with the Kramers-Wannier (KW) duality with $S^1$ topology (see Figure~\ref{fig:g-function1}).
Here, we can use two identities.
One is that the fusion of the KW duality defect and the $+$ boundary is identical to the $0$ boundary.
The other is that the $S^1$ KW duality defect without any operator insertion inside can be replaced by its quantum dimension $\sqrt{2}$.
As a result, we obtain the relation:
\begin{align}
	g_0=\sqrt{2}g_{+}.
\end{align}
One can check that this relation actually holds by constructing the boundary states \cite{Cardy:1989ir}.  However, we emphasize that this relation only depends on the structure of the ``symmetry'' and is independent of the detail of the dynamics, and therefore such relations hold for other theories with the same symmetry as the Ising CFT.
According to the two-dimensional g-theorem,  this formula indicates that the renormalization group flow from the $+$ boundary conditions to the $0$ boundary conditions is prohibited.

In this paper, we consider non-invertible symmetries and boundaries in four dimensions and obtain relations of g-functions in the similar way as in two dimensions.
In particular, we consider various boundary conditions of the four-dimensional $\Zb_2$ lattice gauge theory, and computed the ratios of their g-functions.
We utilize the topological defect associated with the Kramers-Wannier-Wegner (KWW) duality \cite{Wegner:1971app} of this theory, which is obtained in \cite{Koide:2021zxj}.
We employ the lattice approach \cite{Aasen:2016dop,Koide:2021zxj} to describe the topological defects and the boundaries.

\begin{figure}[htbp]
  \centering
  \includegraphics[width=2cm]{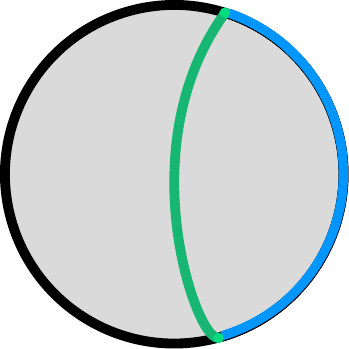}
  \caption{The configuration that we consider in this paper to derive the relations between the g-functions.
The gray disk is a four-dimensional hemisphere on which the $\Zb_2$ gauge theory lives.
The black boundary represents D or \Dtb, and the blue boundary represents N.
The green line represents a KWW duality defect on $D^3$ that has $S^2$ edges on $S^3$ at the boundary of the hemisphere.}
  \label{fig:g-funcIsing}
\end{figure}

Here let us summarize the results.
We consider three boundary conditions.  One is the Dirichlet boundary condition in which all the link variables on the boundary are fixed to be 1.  This boundary condition is denoted by D. Another one is a kind of Dirichlet boundary condition in which all the plaquettes are fixed to be 1.  This boundary condition is denoted by \Dtb.  The other one is the Neumann boundary condition denoted by N.
Let $\gd, \gdt, \gn$ denote the g-functions for D, \Dtb, N, respectively.
We require that the KWW duality defects ending on the boundaries are topological, and obtain the values of the parameters of these boundary conditions.
Although these boundary conditions seem to be related by the fusion of the KWW duality defect, it is not easy to obtain the fusion rule directly in our approach.
Instead, we consider the configuration shown in Figure~\ref{fig:g-funcIsing} to find the relations between the g-functions:
\begin{align}
	\frac{1}{2}\gd =\frac{1}{\sqrt{2}}\gn =\gdt.
\end{align}
This result implies that the renormalization group flow from D to N and from N to \Dtb are prohibited according to the g-theorem in four dimensions. 

Although we only study the $\Zb_2$ lattice gauge theory in this paper, exactly the same non-invertible symmetry exists in other quantum field theories such as the four-dimensional Maxwell theory with $\tau=2i $\cite{Choi:2021kmx} and the $\mathcal{N}=4$ $SU(2)$ super Yang-Mills theory \cite{Kaidi:2021xfk}.
Since our approach is independent of the detail of the dynamics, it is also applicable to these theories.
Moreover, a lot of examples of non-invertible symmetries are found in quantum field theories in four dimensions.  The same approach should be useful to investigate boundaries in these theories.

This paper is organized as follows.
In Section \ref{sec:4dZ2lattice}, we discuss the four-dimensional $\Zb_2$ lattice gauge theory with boundaries and the KWW duality defects connected to the boundaries.
We consider three kinds of boundary conditions related by the fusion of the duality defects.  
We determine the weights of the elements so that the junctions are topological.
Furthermore, we find the expectation value of the duality defects with the $D^3$ topology.
In Section \ref{sec:gfunction}, we find the relations between g-functions by using these results.
Section \ref{sec:discussion} is devoted to conclusions and discussions.

\section{Four-dimensional \texorpdfstring{$\mathbb{Z}_2$}{Z2} lattice gauge theory with boundary and the duality defects}
\label{sec:4dZ2lattice}
\subsection{Four-dimensional \texorpdfstring{$\mathbb{Z}_2$}{Z2} lattice gauge theory}
\label{sec:4dZ2lattice_setup}

In this subsection, we explain the formulation of bulk system.
We consider the 4-dimensional pure $\Zb_2$ lattice gauge theory.
Here, we use the formulation of \cite{Koide:2021zxj}.

We introduce two types of cubic lattices in order to describe the duality defects \cite{Aasen:2016dop,Koide:2021zxj}.
These lattices are dual to each other.
We call the first lattice the active lattice.
We call links and sites in the active lattice active links and active sites, respectively.
In the active lattice, we assign a link variable $U=(-1)^{a}\ (a=0,1)$ to each active link.
We also assign the weights
\begin{align}
s=\frac{1}{\sqrt{2}},\quad
l=\frac{1}{\sqrt{2}}
 \label{bulk weight}
\end{align}
for each active site, and each active link, respectively.
These values are determined in \cite{Koide:2021zxj}.

We call the other lattice the inactive lattice.
We call links and sites in the inactive lattice inactive links and inactive sites, respectively.
The inactive lattice is an auxiliary lattice with no degrees of freedom.
Each weight of inactive sites and links are 1. 
These values are determined in \cite{Koide:2021zxj}.

In order to describe this pair of lattices, we introduce coordinates $(x_1,x_2,x_3,x_4)\in\mathbb{R}^4$.
We define two lattices $\Lambda :=\{(x_1,x_2,x_3,x_4)|x_1,x_2,x_3,x_4\in 2\Zb\}$ and $\hat{\Lambda}:=\{(x_1,x_2,x_3,x_4)|x_1,x_2,x_3,x_4\in 2\mathbb{Z}+1\}$,\footnote{The term ``lattice'' is commonly used to denote a discrete subgroup of the Abelian group $\mathbb{R}^n$. In this regard, $\hat{\Lambda}$ does not qualify as a lattice. Nevertheless, we refer to $\hat{\Lambda}$ as a lattice due to its congruence with a traditional lattice.} which are dual to each other.
We do not fix the roles of $\Lambda$ and $\hat{\Lambda}$ for convenience.
In some cases, $\Lambda$ is the active lattice and $\hat{\Lambda}$ is the inactive lattice.  In the other cases, $\hat{\Lambda}$ is the active lattice and $\Lambda$ is the inactive lattice.

Let $U_i=(-1)^{a_i}\ (i=1,2,3,4)$ be the link variables of the four links in a plaquette and $K$ be a real parameter.
We also call $a_i$ a link variable.
We assign the Boltzmann weight to this plaquette:
\begin{align}
	W(a_1,a_2,a_3.a_4)=\exp\left(K(-1)^{(a_1+a_2+a_3+a_4)}\right).
  \label{Boltzmann weight}
\end{align}
The partition function of the $\Zb_2$ lattice gauge theory is given by
\begin{align}
    Z=\sum_{\{a\}}\left(\prod_{\substack{ \text{active}\\ \text{sites}}}s\right)\left(\prod_{\substack{ \text{active}\\ \text{links}}}l\right)\prod_{i\in C}W(a_{j_1(i)},a_{j_2(i)},a_{j_3(i)},a_{j_4(i)}),
    \label{partition function}
\end{align}
where $C$ is the set of all active plaquettes and $j_1(i),j_2(i),j_3(i),j_4(i)$ are the four active links in the active plaquette $i$.  $a_j$ is the link variable of the link $j$.

We fix the parameter $K$ to the self-dual point: 
\begin{align}
	K_c=-\frac{1}{\sqrt{2}}\log(-1+\sqrt{2}).
  \label{Kc}
\end{align}
At this self-dual point, we can construct non-invertible KWW duality defect.
A building block of the KWW duality defects is a tetrahedral prism that contains two tetrahedrons.  Each tetrahedron in a building block includes an active link and an inactive link.
We assign the weight $D(a,\tilde{a})$ to a building block, where $a$ and $\tilde{a}$ are link variables of the two active links in the building block.
$D(a,\tilde{a})$ is determined in \cite{Koide:2021zxj} so that the duality defect is topological:
\begin{align}
	D(a,\tilde a)=(-1)^{a\tilde a}.
  \label{Da}
\end{align}

\subsection{Boundary conditions}
\label{sec:boundary conditions}
In this subsection, we introduce three types of boundary conditions.
We consider one Neumann boundary condition and two types of Dirichlet boundary conditions for the link variable $a$.

We consider spacetime $M$:
\begin{align}
    M:=\{(x_1,x_2,x_3,x_4)|x_1\geq 0\}.
  \label{spacetime}
\end{align}
The boundary of $M$ is located at  $x_1=0$ and is a part of $\Lambda$.  
We consider both cases that $\Lambda$ is the inactive lattice and the active lattice.

First, we consider the case that $\Lambda$ is the inactive lattice.
In this case, there are no active links on the boundary.
Thus, we do not impose any condition on the active links, and therefore this boundary condition is the free or Neumann boundary condition.
We denote this Neumann boundary condition by N.
We can assign arbitrary weights for each link and site on the boundary.
Here, we fix these values to 1 for simplicity.
Then, the partition function $\Z{N}$ with this boundary condition is given by
\begin{align}
    \Z{N}=\sum_{\{a\}}\left(\prod_{\substack{ \text{active}\\ \text{sites}}}s\right)\left(\prod_{\substack{ \text{active}\\ \text{links}}}l\right)\prod_{i\in C}W(a_{j_1(i)},a_{j_2(i)},a_{j_3(i)},a_{j_4(i)}).
\end{align}
Now, we use the same notation as \eqref{partition function}.

Second, we consider the case that $\Lambda$ is the active lattice.
We study two types of Dirichlet boundary conditions.
These two boundary conditions look quite similar, but actually they are different.

One Dirichlet boundary condition is that all the link variables $a$ on the boundary are fixed to $0$.
We call this boundary condition D.
The weights of a site, a link and, a plaquette at the boundary are denoted by $\bsD$, $\blD$, and $\bWD$, respectively.
We will determine these values later.
The partition function $\Z{D}$ with this boundary condition is given by
\begin{align}
  \Z{D}=\sum_{\{a\}}\left(\prod_{\substack{ \text{bulk active}\\ \text{sites}}}s\right)\left(\prod_{\substack{ \text{bulk active}\\ \text{links}}}l\right)\left(\prod_{\substack{ \text{boundary}\\ \text{sites}}}\bsD\right)\left(\prod_{\substack{ \text{boundary}\\ \text{links}}}\blD\right)
  \left(\prod_{\substack{ \text{boundary}\\ \text{plaquettes}}}\bWD\right)\prod_{i\in C}W(a_{j_1(i)},a_{j_2(i)},a_{j_3(i)},a_{j_4(i)}).
\end{align}
Here, $C$ is the set of all the active plaquettes which contain at least one bulk active link, and $j_1(i),j_2(i),j_3(i),j_4(i)$ are active links in active plaquette $i$.
The summation for $\{a\}$ is taken for all possible configuration which satisfy the boundary condition.

Another Dirichlet boundary condition is that all plaquettes on the boundary are 0.
We call this boundary condition \Dtb.
The weights for a site and a link at the boundary are denoted by $\bsDt$ and $\blDt$, respectively. The boundary Boltzmann weight is denoted by $\bWDt\deltamod{a_1+a_2+a_3+a_4}$, where $a_i\ (i=1,2,3,4)$ are the link variables of the links in this plaquette.
Here, $\deltamod{a}$ is defined as
\begin{align}
	\deltamod{a}:=
	 \begin{cases}
	  0 & (a:\text{odd})\\
	  1 & (a:\text{even})
	 \end{cases}.
\end{align}
The partition function $\Z{\Dtb}$ with this boundary condition is as follows:
\begin{align}
  \Z{\Dtb}=\sum_{\{a\}}&\left(\prod_{\substack{ \text{bulk active}\\ \text{sites}}}s\right)\left(\prod_{\substack{ \text{bulk active}\\ \text{links}}}l\right)
  \left(\prod_{\substack{\text{boundary}\\ \text{sites}}}\bsDt\right)\left(\prod_{\substack{ \text{boundary}\\ \text{links}}}\blDt\right)\nonumber\\
  &\times\left(\prod_{k\in C_B}\bWDt\deltamod{a_{j_1(k)}+a_{j_2(k)}+a_{j_3(k)}+a_{j_4(k)}}\right)\prod_{i\in C}W(a_{j_1(i)},a_{j_2(i)},a_{j_3(i)},a_{j_4(i)}).
\end{align}
Now,$C_B$ is the set of all boundary plaquettes, $C$ is the set of all bulk plaquettes, and $j_1(i),j_2(i),j_3(i),j_4(i)$ are links in the plaquette $i$.
The summation $\{a\}$ is taken for all possible configurations of the link variables. The boundary condition is imposed by the boundary Boltzmann weight.

\begin{figure}[htbp]
\centering
\includegraphics[width=2cm]{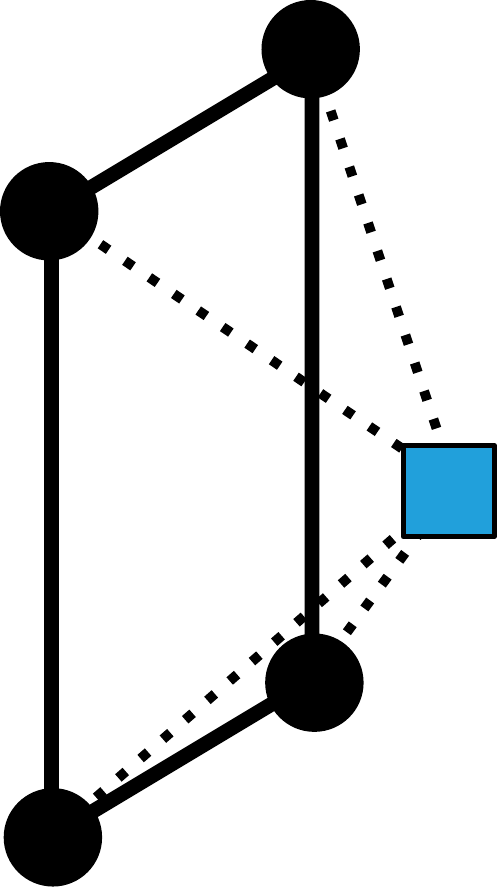}
\caption{
  A schematic picture of a \bbb.
  The black plaquette represents a plaquette on the boundary and the blue square dot represents a link in the bulk.}
\label{fig:q16}
\end{figure}
There are two kinds of basic units on the boundary.
One of them is the convex hull of a plaquette on the boundary lattice $\Lambda$ and the closest link to the plaquette in the bulk in $\hat{\Lambda}$ as shown in Figure~\ref{fig:q16}.
We call this basic unit \bbb.
The surface of a \bbb contains six cells; four of them are tetrahedrons and two of them are square pyramids.
For example, a \bbb is the convex hull of the six points $(0,0,0,0)$, $(0,0,0,2)$, $(0,0,2,0)$, $(0,0,2,2)$, $(1,-1,1,1)$ and $(1,1,1,1)$.

\begin{figure}[htbp]
\centering
\includegraphics[width=4cm]{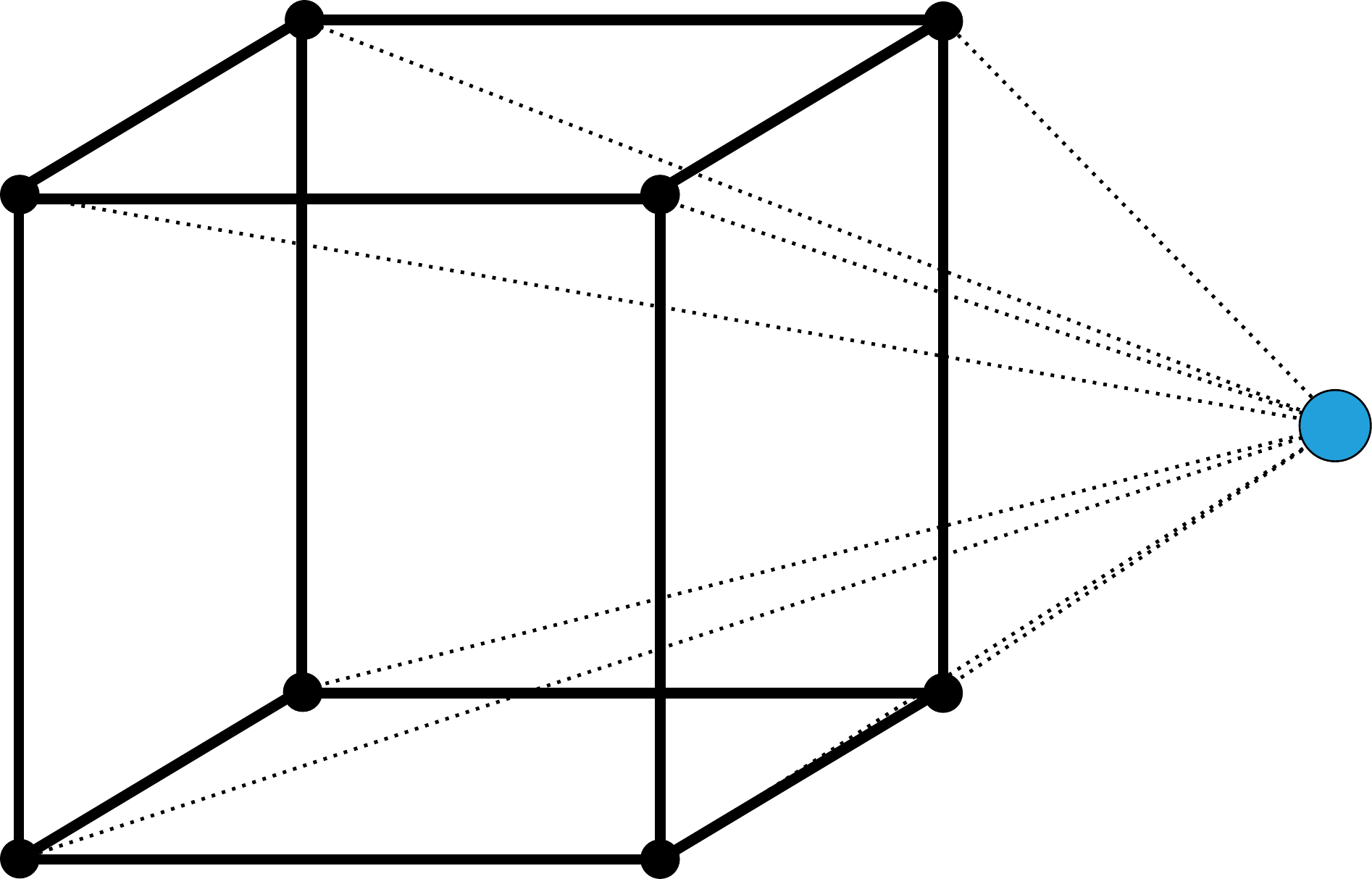}
\caption{A schematic picture of a \cubebb.
  The black dots and lines represent sites and links on the boundary, respectively. 
  The blue dot represents a site in the bulk.}
      \label{fig:bcube}
\end{figure}

The other basic unit is the convex hull of a three-dimensional cube on the boundary and a site closest to the cube in the bulk as shown in Figure~\ref{fig:bcube}.
We call this unit a \cubebb.
The surface of a \cubebb contains seven cells; six of them are square pyramids and one of them is a cube.
We consider the commutation relation on a \cubebb in Sec.~\ref{sec:topological defects on boundary}.

\subsection{Topological defects ending on the boundary}
\label{sec:topological defects on boundary}
Let us consider the KWW duality defects ending on the boundary.
The KWW duality swap the active lattice and the inactive lattice.
Therefore, we expect that the KWW duality defects connect N and D or \Dtb on the boundary.
In this subsection, we study conditions that the KWW duality defects are topological on the boundary; we call this conditions ``boundary defect commutation relations.''
By the boundary defect commutation relations, we decide the boundary Boltzmann weights and the weights of the sites and the links on the boundary.

\begin{figure}[htbp]
  \centering
\includegraphics[width=3cm]{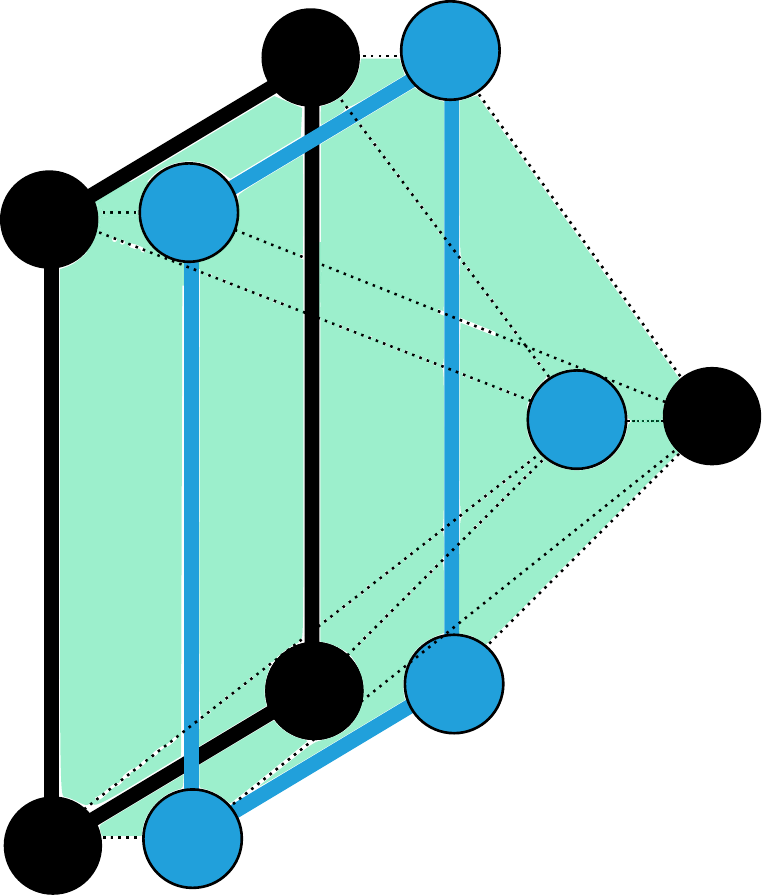}
\caption{
 A schematic picture of a building block of KWW duality defects on a boundary.
 The unit is defined on a doubled square pyramid.
 Each square pyramid includes a boundary plaquette and a bulk site closest to the plaquette.}
\label{fig:Pyra-unit}
\end{figure}

Here, we introduce a building block of KWW duality defects that end on the boundary in addition to the tetrahedral prism in the bulk considered in \cite{Koide:2021zxj}.
This additional building block of KWW duality defects is a doubled square pyramid that has a pair of active and inactive plaquette on the boundary and a pair of active and inactive sites closest to the plaquette in a bulk as shown in Figure~\ref{fig:Pyra-unit}.

We consider two types of boundary defect commutation relations; one is associated with a \cubebb, and the other is associated with a \bbb.
By combining defect commutation relations studied in \cite{Koide:2021zxj} and boundary defect commutation relations on a \cubebb and a \bbb, KWW duality defects can be deformed smoothly even when KWW duality defects end on the boundary.
We obtain the weights of the elements on the boundary from the defect commutation relations on a \cubebb.
We study the boundary defect commutation relations on a \cubebb in this section.
On the other hand, the boundary defect commutation relations on a \bbb are satisfied for arbitrary values of the weights of the elements on the boundary.
We study boundary defect commutation relations on a \bbb in Appendix~\ref{sec:commutation relations on bbb}.

\begin{figure}[htbp]
  \begin{tabular}{ccc}
    \begin{minipage}[t]{0.3\hsize}
      \centering
      \includegraphics[width=4cm]{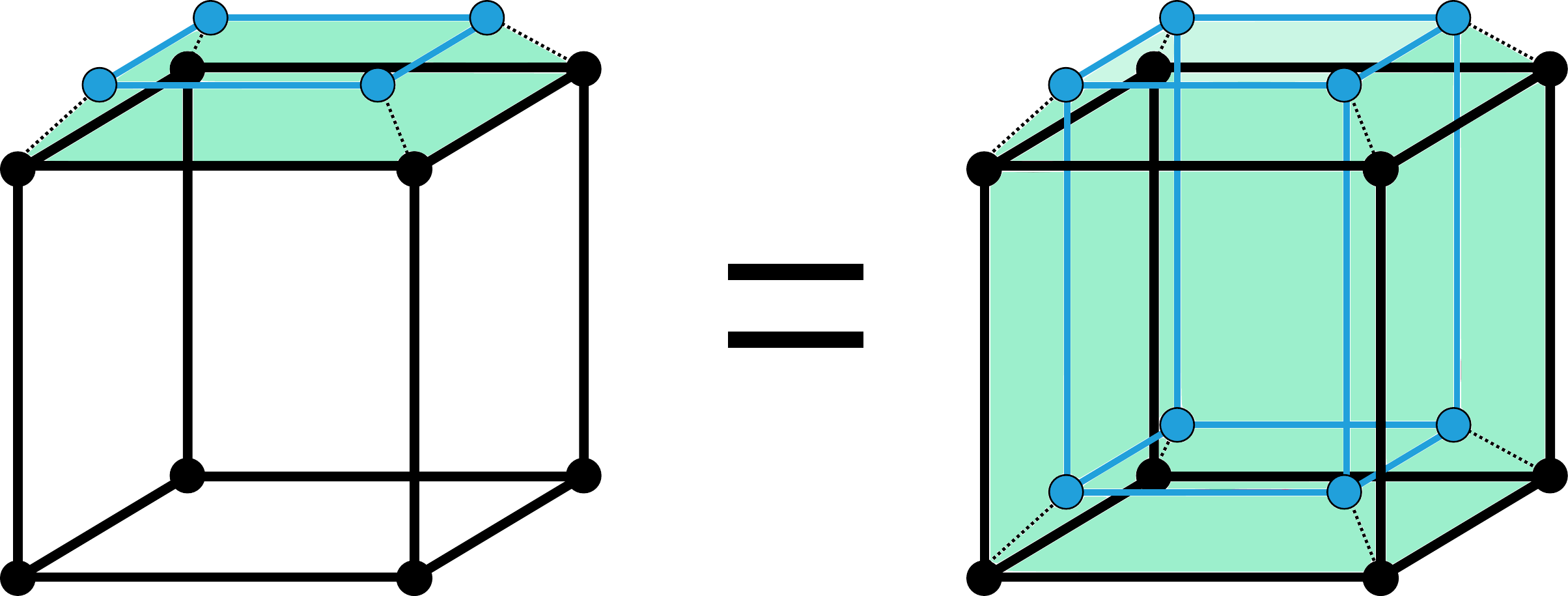}
      \subcaption{}
      \label{PPP1-5}
    \end{minipage} &
    \begin{minipage}[t]{0.3\hsize}
      \centering
      \includegraphics[width=4cm]{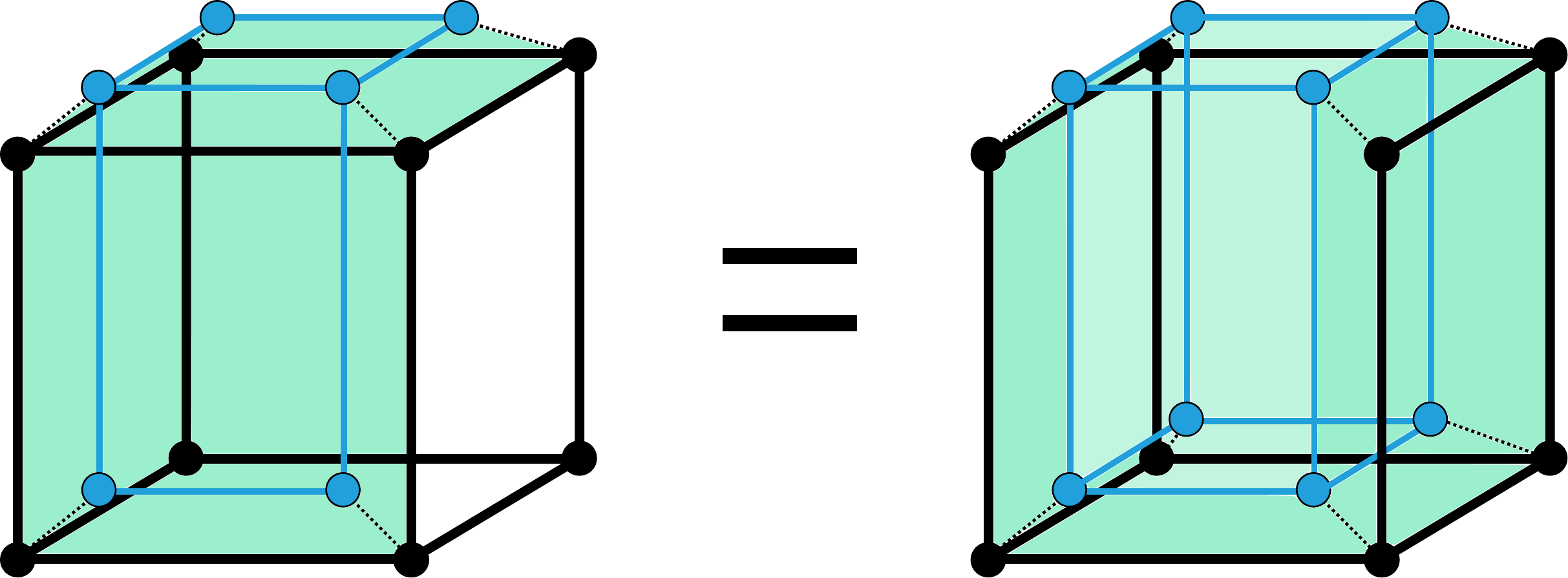}
      \subcaption{}
      \label{PPP2-4}
    \end{minipage} &
    \begin{minipage}[t]{0.3\hsize}
      \centering
      \includegraphics[width=4cm]{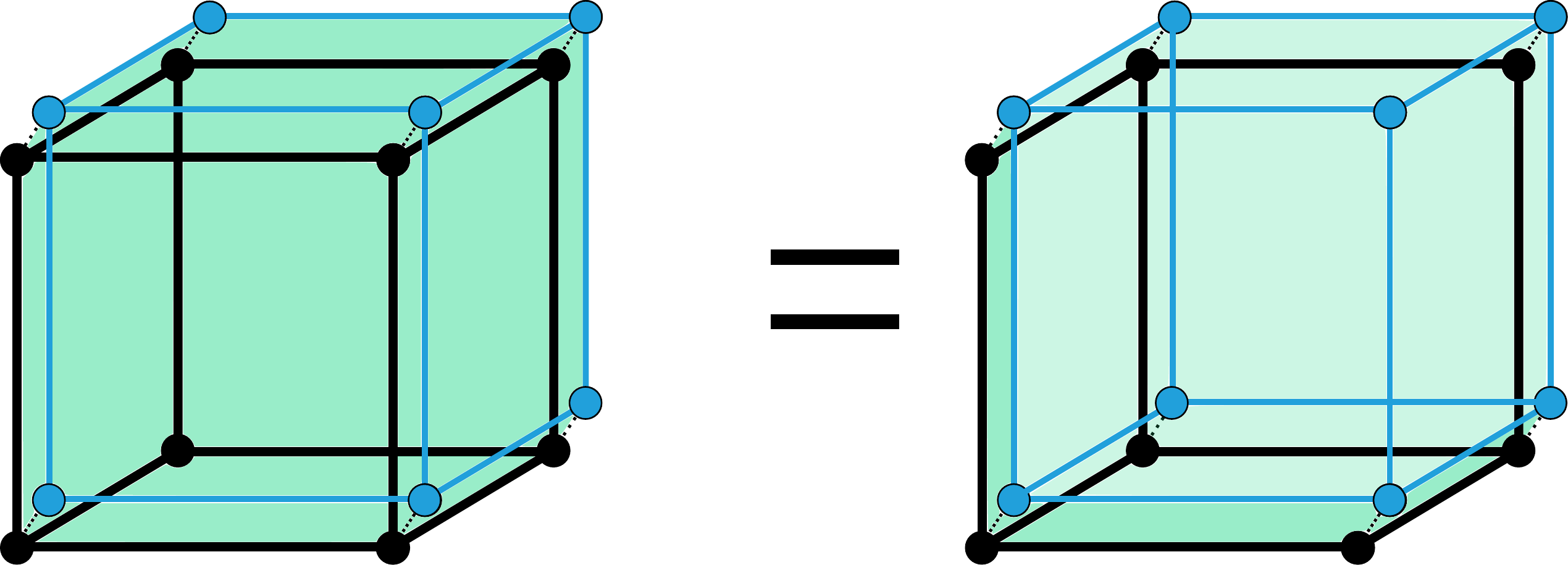}
      \subcaption{}
      \label{PPP3-3A}
    \end{minipage} \\
    \begin{minipage}[t]{0.3\hsize}
      \centering
      \includegraphics[width=4cm]{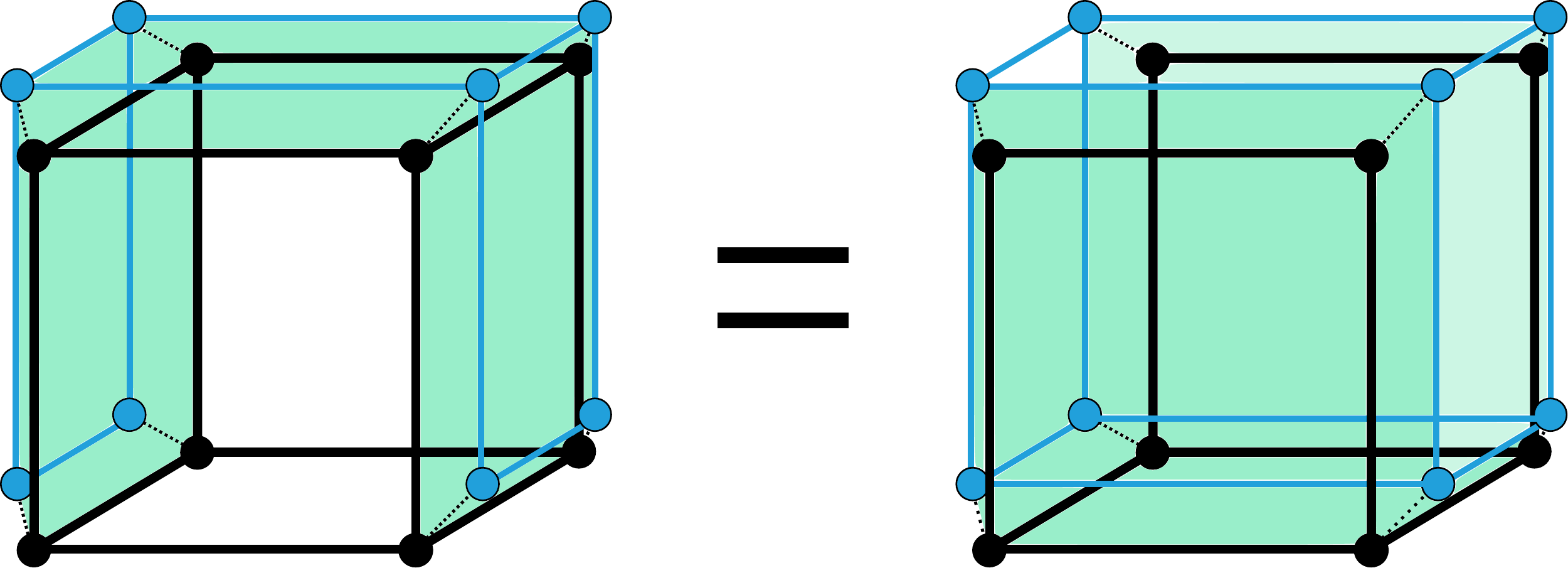}
      \subcaption{}
      \label{PPP3-3B}
    \end{minipage} &
  \begin{minipage}[t]{0.3\hsize}
    \centering
    \includegraphics[width=4cm]{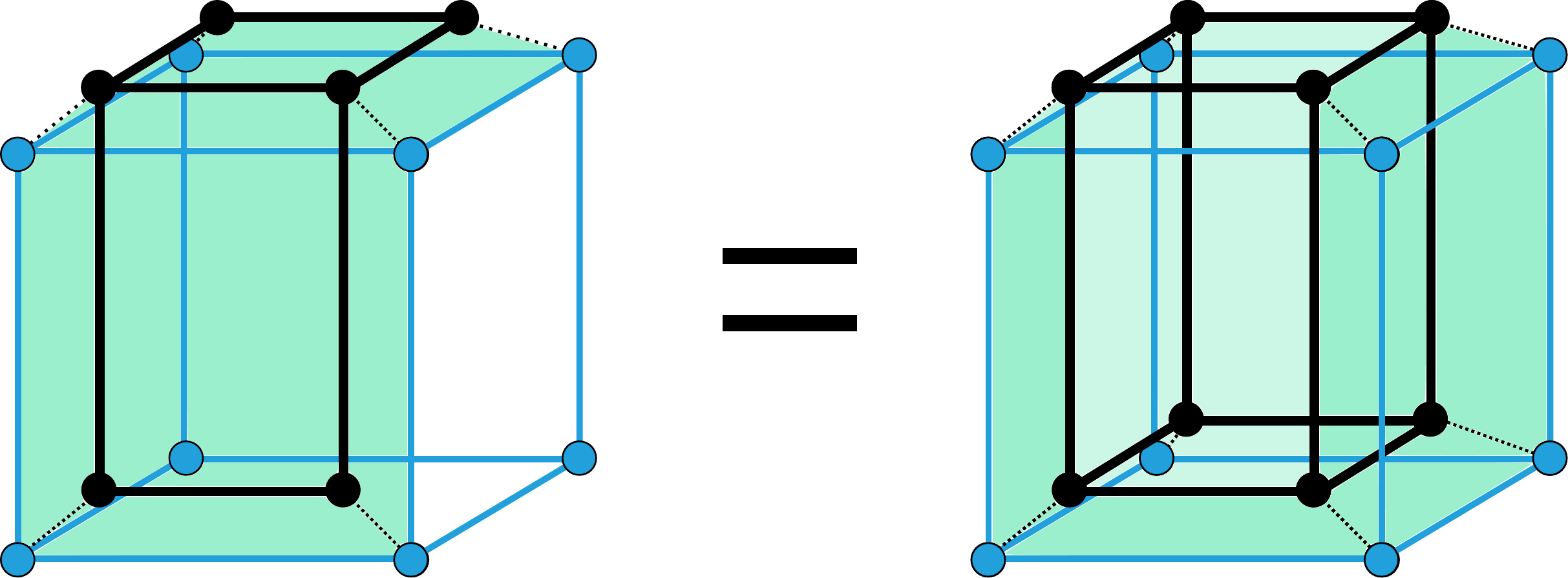}
    \subcaption{}
    \label{PPP4-2}
  \end{minipage} &
  \begin{minipage}[t]{0.3\hsize}
    \centering
    \includegraphics[width=4cm]{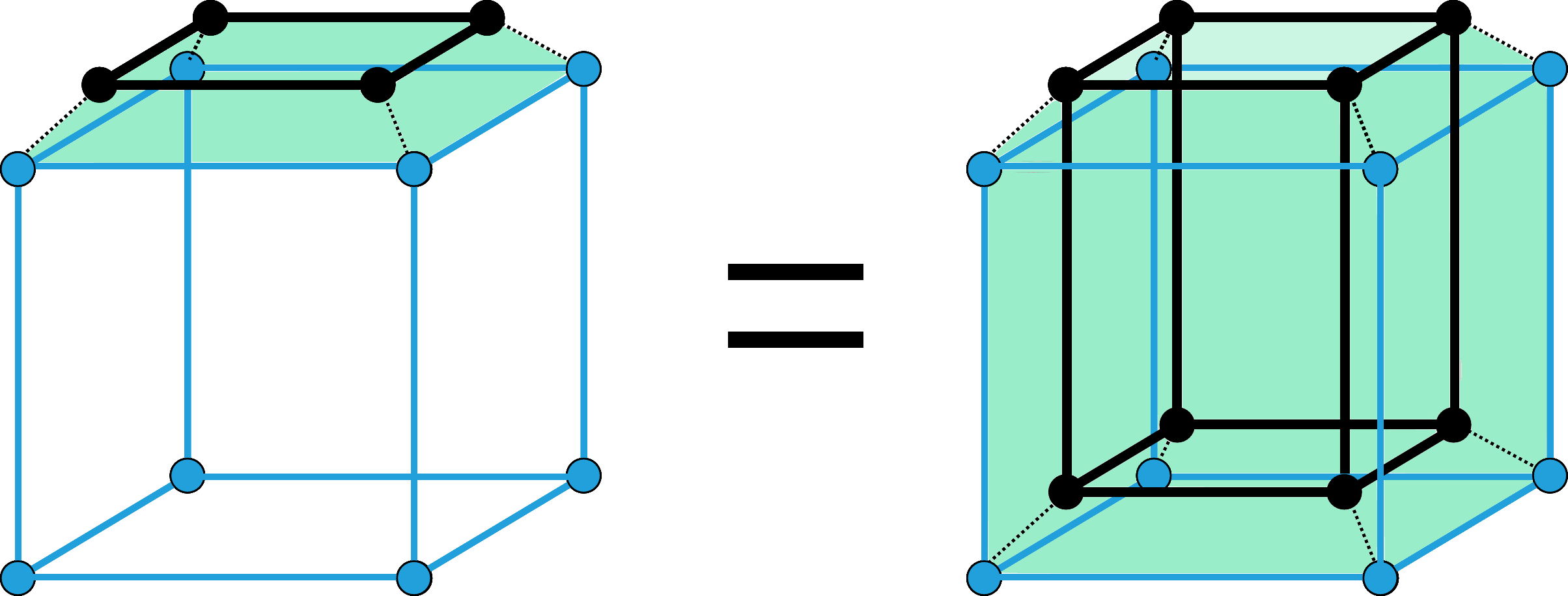}
    \subcaption{}
    \label{PPP5-1}
  \end{minipage}
\end{tabular}
   \caption{Schematic pictures of boundary defect commutation relations on a \cubebb. Blue lattices represent the N boundary condition. Black lattices represent D or \Dtb boundary condition. Some of square pyramids are filled by the KWW duality defects which are represented by green surfaces.}
   \label{fig:boundery defect commutation relations on a cubebb}
\end{figure}

Let us consider boundary defect commutation relations associated with a \cubebb.
A \cubebb contains six square pyramids on which KWW defects can be placed.  A defect commutation relation relates a configuration of a KWW duality defect and another one with the same topology that are different only around a \cubebb.
There are six boundary defect commutation relations of the KWW defect connecting D and N up to rotation as shown in Figure~\ref{fig:boundery defect commutation relations on a cubebb}.  There are also six boundary defect commutation relations connecting \Dtb and N depicted by the same figures.

The defect commutation relations associated with D and N boundary conditions depicted by Figures \ref{fig:boundery defect commutation relations on a cubebb}(\subref{PPP1-5})--(\subref{PPP5-1}) are given, respectively, by
\begin{align}\bWD^{6}\blD^{12}\bsD^{8}\pD&=\bWD^{5}\blD^{12}\bsD^{8}\pD^{5}, \label{cube 1-5 on D}\\
  \bWD^{6}\blD^{12}\bsD^{8}\pD^{2}&=\bWD^{4}\blD^{11}\bsD^{8}\pD^{4}, \label{cube 2-4 on D}\\
  \bWD^{6}\blD^{12}\bsD^{8}\pD^{3}&=\bWD^{3}\blD^{9}\bsD^{7}\pD^{3},  \label{cube 3-3A on D}\\
  \bWD^{6}\blD^{12}\bsD^{8}\pD^{3}&=\bWD^{3}\blD^{10}\bsD^{8}\pD^{3}, \label{cube 3-3B on D}\\
  \bWD^{2}\blD^{7}\bsD^{6}\pD^{2}&=\bWD^{6}\blD^{12}\bsD^{8}\pD^{4}, \label{cube 4-2 on D}\\
\bWD\blD^{4}\bsD^{4}\pD&=\bWD^{6}\blD^{12}\bsD^{8}\pD^{5}. \label{cube 5-1 on D}
\end{align}
Note that the Boltzmann weight of the top active plaquette in the right-hand side of Figure~\ref{fig:boundery defect commutation relations on a cubebb}(\subref{PPP1-5}) does not contribute to the partition function as shown in Eq.~(\ref{cube 1-5 on D}).
This is because the inside of this plaquette does not belong to D or \Dtb, but it does belong to N.
Eqs.~\eqref{cube 1-5 on D}--\eqref{cube 5-1 on D} are not independent.  They are equivalent to the three equations:
\begin{align}
  \bWD=\pD^4, \label{rel wdp}\\
  \bWD^3\blD^2=1,\label{rel wdl} \\
  \bWD^3\blD^3\bsD=1.\label{rel wdls}
\end{align}

On the other hand, the defect commutation relations associated with \Dtb and N boundary conditions depicted by Figures \ref{fig:boundery defect commutation relations on a cubebb}(\subref{PPP1-5})--(\subref{PPP5-1}) are given, respectively, by
\begin{align}
\bWDt^{6}\blDt^{12}\bsDt^{8}\pDt&=\bWDt^{5}\blDt^{12}\bsDt^{8}\pDt^{5}, \label{cube 1-5 on Dt}\\
  2\bWDt^{6}\blDt^{12}\bsDt^{8}\pDt^{2}&=\bWDt^{4}\blDt^{11}\bsDt^{8}\pDt^{4}, \label{cube 2-4 on Dt}\\
  2^3\bWDt^{6}\blDt^{12}\bsDt^{8}\pDt^{3}&=\bWDt^{3}\blDt^{9}\bsDt^{7}\pDt^{3},  \label{cube 3-3A on Dt}\\
  2^2\bWDt^{6}\blDt^{12}\bsDt^{8}\pDt^{3}&=\bWDt^{3}\blDt^{10}\bsDt^{8}\pDt^{3}, \label{cube 3-3B on Dt}\\
  \bWD^{2}\blD^{7}\bsD^{6}\pD^{2}&=2^5\bWD^{6}\blD^{12}\bsD^{8}\pD^{4}, \label{cube 4-2 on Dt}\\
\bWD^{1}\blD^{4}\bsD^{4}\pD&=2^8\bWD^{6}\blD^{12}\bsD^{8}\pD^{5}. \label{cube 5-1 on Dt}
\end{align}
Here, $\delta^{\mathrm{mod}\ 2}$ and summations of link variables common to both sides are omitted.
The numerical coefficients $2^n$ in Eqs.~\eqref{cube 2-4 on Dt}--\eqref{cube 5-1 on Dt} are the numbers of the configurations of the link variables that satisfy the boundary condition.
Eqs.~\eqref{cube 1-5 on Dt}--\eqref{cube 5-1 on Dt} are not independent.  They are equivalent to the three equations:
\begin{align}
  \bWDt=\pDt^4, \label{rel wdtp}\\
  \bWDt^3\blDt^2=1,\label{rel wdtl}\\
  2\bWDt^3\blDt^3\bsDt=1. \label{rel wdtls}
\end{align}

\subsection{\texorpdfstring{\diskvs}{D3 expectation values}}
\label{sec:D3expectation value}

\begin{figure}[htbp]
\centering
\includegraphics[width=7cm]{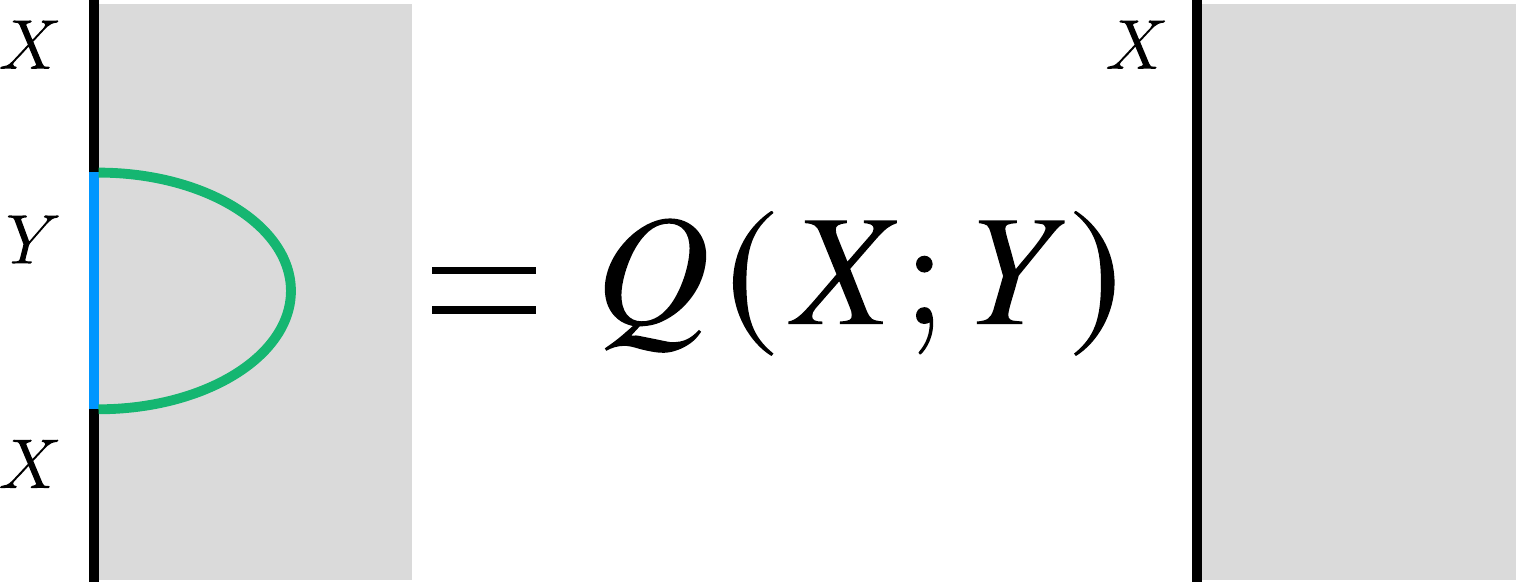}
\caption{Derivation of the \diskv.
The gray region represents the bulk on which the $\Zb_2$ gauge theory lives.
The black or blue vertical lines represent boundaries. 
The green line represents the KWW duality defect on $D^3$ whose edge is $S^2$ on the boundary. 
The black lines represent the $X$ boundary condition. 
The blue line inside the $S^2$ represents the $Y$ boundary. When no other operator is contained inside the $D^3$, we can replace it with \diskv $\Q{$X$}{$Y$}$.}
\label{fig:Qcost}
\end{figure}
\begin{figure}[htbp]
  \begin{minipage}[b]{0.45\linewidth}
    \centering
    \includegraphics[width=6cm]{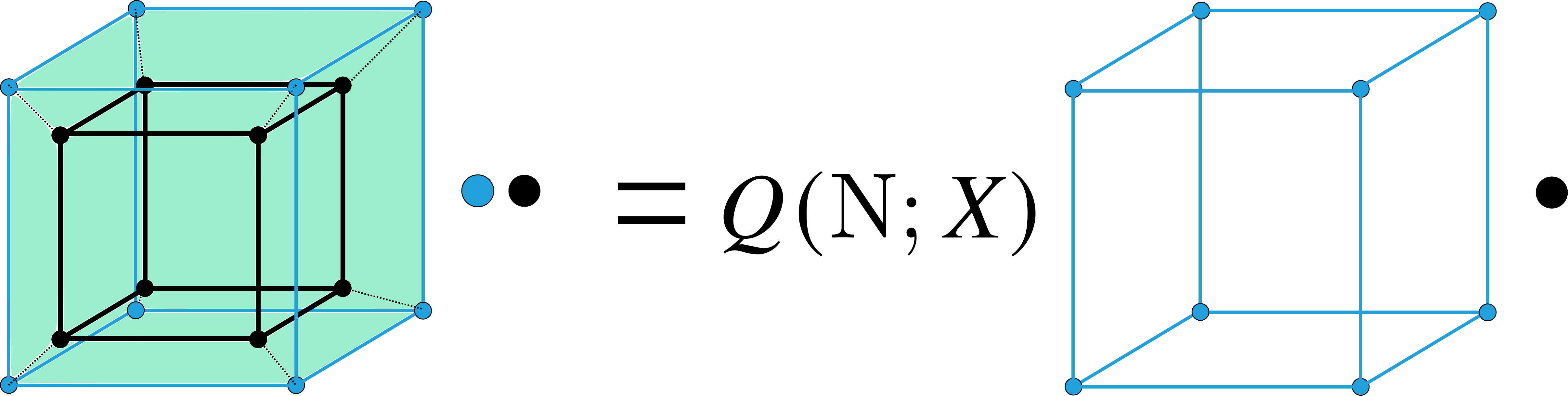}
    \subcaption{}
    \label{fig:P6-0W}
  \end{minipage}
  \begin{minipage}[b]{0.45\linewidth}
    \centering
    \includegraphics[width=6cm]{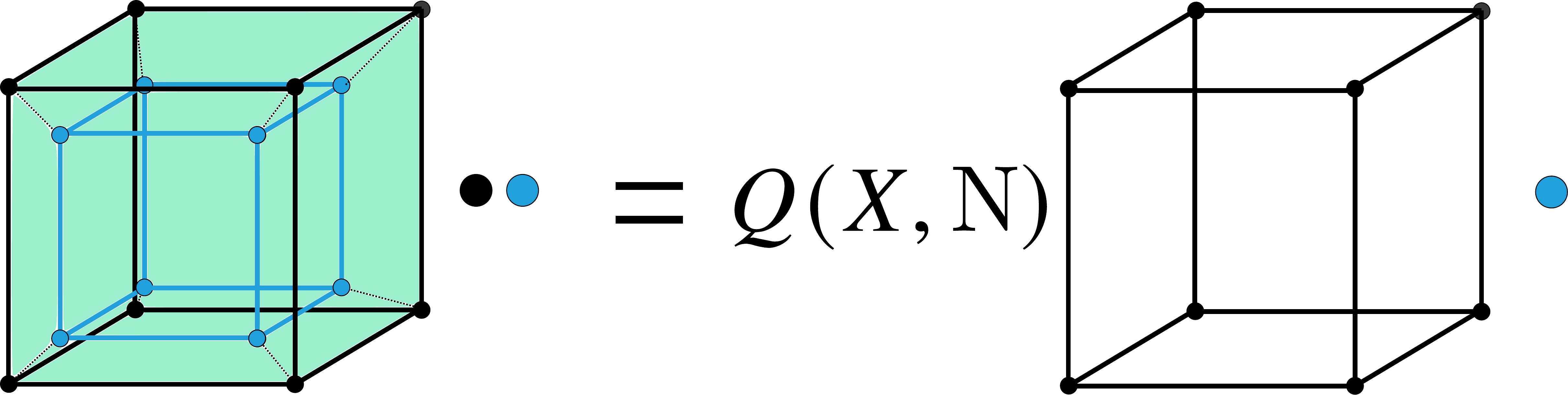}
    \subcaption{}
    \label{fig:P6-0B}
  \end{minipage}
  \caption{Configrations used to obtain \diskvs.
The KWW duality defect connects N boundary condition and $X$ boundary condition, where $X$ is D or \Dtb.
In the left-hand side, all the square pyramids on the surface of a \cubebb are filled by the KWW duality defect.
On the other hand, there is no KWW duality defect in the right-hand side.
The roles of $X$ and N are interchanged between (a) and (b).}
\label{fig:P6-0}
\end{figure}

We consider a KWW duality defect on $D^3$ whose edge is $S^2$ on the boundary.  See Figure \ref{fig:Qcost}.
On the boundary, the inside of this $S^2$ is $Y$ boundary and the outside of it is $X$ boundary, where $Y$ is N and $X$ is D or \Dtb, or vice versa.
When no other operator is contained inside this $D^3$, we can replace it with a topological local operator on the boundary.
This topological local operator turn out to be a c-number times the identity operator.
Let us call this c-number the ``\diskv'' and denote it by $\Q{$X$}{$Y$}$.
In this section, we compute the \diskvs.
In the following, we determine the \diskvs by considering the duality defects placed on all the square pyramids of a \cubebb as in the left-hand side of Figure \ref{fig:P6-0}.

First, we consider $\Q{N}{D}$ and $\Q{D}{N}$.  The relations of Figure \ref{fig:P6-0} reads
\begin{align}
  \bWD^{6}\blD^{12}\bsD^{8}\pD^{6}&=\Q{N}{D},\label{Qrel N-D}\\
  \bWD^{6}\blD^{12}\bsD^{8}\pD^{6}s&=\Q{D}{N}\bWD^{6}\blD^{12}\bsD^{8}.\label{Qrel D-N}      
\end{align}
Here, $s=\frac{1}{\sqrt{2}}$ in \eqref{Qrel D-N} is the weight assigned to the active site in the bulk \eqref{bulk weight}.
By using Eqs.~\eqref{rel wdp}, \eqref{rel wdl}, \eqref{rel wdls}, 
$\Q{N}{D}$ and $\Q{D}{N}$ are expressed in terms of $\bWD$ as
\begin{align}
	\Q{N}{D}&=\bWD^{\frac{3}{2}},\label{Qexp N-D}\\
	\Q{D}{N}&=\frac{1}{\sqrt{2}}\bWD^{\frac{3}{2}}.\label{Qexp D-N}
\end{align}

Next, we consider $\Q{N}{\Dtb}$ and $\Q{\Dtb}{N}$.  The relations of Figure \ref{fig:P6-0} reads
\begin{align}
  2^{7}\bWDt^{6}\blDt^{12}\bsDt^{8}\pDt^{6}&=\Q{N}{\Dtb}, \label{Qrel N-Dt}\\
  \bWDt^{6}\blDt^{12}\bsDt^{8}\pDt^{6}s&=\Q{\Dtb}{N} \bWDt^{6}\blDt^{12}\bsDt^{8}. \label{Qrel Dt-N}
\end{align}
Here, the coefficient $2^7$ on the left side of Eq.~\eqref{Qrel N-Dt} is the number of possible configurations of the boundary link variables that satisfy the boundary condition.
The \diskvs $\Q{N}{\Dtb}$, $\Q{\Dtb}{N}$ are expressed in terms of $\bWDt$ by using the relations \eqref{rel wdtp}, \eqref{rel wdtl}, \eqref{rel wdtls} as
\begin{align}
	\Q{N}{\Dtb}&=\frac{1}{2}\bWDt^{\frac{3}{2}}, \label{Qexp N-Dt}\\
	\Q{\Dtb}{N}&=\frac{1}{\sqrt{2}}\bWDt^{\frac{3}{2}}.\label{Qexp Dt-N}
\end{align}

By the above calculations, we find that a duality defect with $D^3$ topology with no other operator inside can be replaced by the \diskv times the identity operator.
We also express the \diskvs in terms of the Boltzmann weight on the boundary.

\section{Relations between g-functions}
\label{sec:gfunction}
In this section, we derive the relations between g-functions. Since it is not easy to obtain the fusion rule directly in the Aasen, Mong, and Fendley (AMF) approach~\cite{Aasen:2016dop,Aasen:2020jwb}, we obtain those relations from the \diskvs $\Q{$X$}{$Y$}$ that we obtained in Section \ref{sec:D3expectation value}.

\begin{figure}[htbp]
  \centering
  \includegraphics[width=10cm]{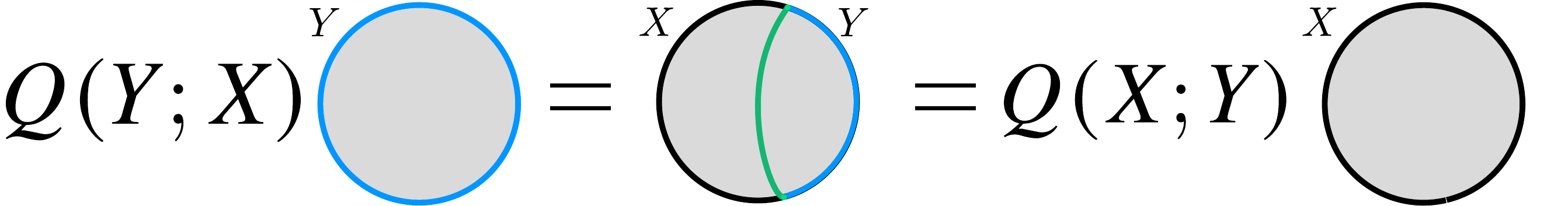}
  \caption{The derivation of the relation between g-functions.
  The green line represents the KWW duality defect on $D^3$ ending on $S^2$ on the boundary of the four-dimensional hemisphere on which the $\Zb_2$ lattice gauge theory lives. This KWW duality defect connects the boundary conditions $X$ and $Y$.The black boundary represents the boundary condition $X$, and the blue boundary represents the boundary condition $Y$.
  We use the identity in Figure \ref{fig:Qcost} in two different ways: the left-hand side and the right-hand side. 
  }
  \label{fig:g-func}
\end{figure}

We consider the $\Zb_2$ lattice gauge theory on a four-dimensional hemisphere.
Then, we place a duality defect on $D^3$ that ends on $S^2$ on the boundary of the four-dimensional hemisphere.
The boundary conditions of the theory change from D to N or from \Dtb to N at the edge of this KWW duality defect.
Let $\langle V\rangle$ be the partition function of this system.
As we found in Section \ref{sec:D3expectation value}, this defect can be replaced by the \diskv.
In particular, since this spacetime is a four-dimensional hemisphere, we can use this identity in two different ways as shown in Figure \ref{fig:g-func}.  As a result, we obtain the relation:
\begin{align}
	\g{X} Q(X;Y)=\langle V\rangle=\g{Y} Q(Y;X),
	\label{eq:g-func}
\end{align}
where $X$ and $Y$ denote the two boundary conditions connected by the duality defect.
In particular, Eq.~\eqref{eq:g-func} for $(X,Y)=(\mbox{D},\mbox{N})$ and $(X,Y)=(\mbox{\Dtb},\mbox{N})$ read
\begin{align}
	\gd\Q{D}{N} =\gn \Q{N}{D},
	\\
	\gdt \Q{\Dtb}{N} =\gn \Q{N}{\Dtb},
\end{align}
where $\gd$, $\gdt$, and $\gn$ denote the g-functions of D, \Dtb, and N, respectively.
From these equations and the expressions of the \diskvs \eqref{Qexp N-D}, \eqref{Qexp D-N}, \eqref{Qexp N-Dt}, \eqref{Qexp Dt-N}, we obtain the relations between g-functions:
\begin{align}
	\frac{1}{2}\gd =\frac{1}{\sqrt{2}}\gn =\gdt. \label{relations_g-functions}
\end{align}

Due to the four-dimensional g-theorem, the g-function is monotonically increasing along the boundary renormalization group flow.  Therefore, the relation \eqref{relations_g-functions} implies that the boundary renormalization group flows from \Dtb to N and from N to D are prohibited.

Although, it is not easy to obtain the fusion rules in the AMF approach, we obtain them indirectly from Eq.~\eqref{relations_g-functions} and the $S^3$ expectation value of the duality defect $\frac{1}{\sqrt 2}$:
\begin{align}
  \mbox{D}\times K = \mbox{N},\quad
  \mbox{N}\times K =\mbox{\Dtb},
  \label{boundary fusion rule}
\end{align}
where $K$ is the KWW duality defect.
This is also consistent with the bulk fusion rule $K\times K=C$, where $C$ is the codimension one condensation defect of the $\Zb_2$ one-form symmetry \cite{Roumpedakis:2022aik,Choi:2022zal}; the fusion rule $\mbox{D}\times C=\mbox{\Dtb}$ derived from this bulk fusion rule and Eq.~\eqref{boundary fusion rule} agrees with the definitions of D and \Dtb.

\section{Conclusion and discussion}
\label{sec:discussion}
In this paper, we study four-dimensional $\Zb_2$ lattice gauge theory with three types of boundary conditions: D, \Dtb, and N.
We determine the weights of the elements on the boundary
so that the KWW duality defects can be smoothly deformed.
With these solutions, we get the ratios of the hemisphere partition functions with N, D and \Dtb.
These ratios constrain possible boundary renormalization group flows.

Let us explain the boundary conditions N, D and \Dtb in the context of the continuum limit.  We will consider the low energy limit in the deconfinement phase $K>K_c$ instead of the critical point.  The low energy theory is the topological $\Zb_2$ gauge theory, which is described, for example, by the BF theory:
\begin{align}
  S_{\text{BF}}=\frac{2i}{2\pi}\int BdA,
\end{align}
where $A$ and $B$ are 1-form and 2-form U(1) gauge fields, respectively.  In this context, D boundary condition flows to the Dirichlet boundary condition $A|_{\text{boundary}}=0$, while both N and \Dtb boundary conditions flow to the Neumann boundary condition $B|_{\text{boundary}}=0$.  These boundary conditions of the topological $\Zb_2$ gauge theory are referred to as the ``Higgsed'' boundary and the ``deconfined'' boundary, respectively in \cite{Ji:2022iva}.  Another boundary condition known as the ``twisted'' boundary condition, is also explored in \cite{Ji:2022iva}.  Uncovering the microscopic description of the twisted boundary in our $\Zb_2$ lattice gauge theory and its behavior in the KWW duality presents an intriguing problem for future investigation.

In this paper, we only investigate the commutation relations of the KWW defects ending on the flat boundary;
this is enough for determining the ratios of g-functions.
However, we have to determine the weights of the elements on the corners in order to define the partition function on a spacetime that has hemisphere topology, e.g. a hyper cube.
This is an interesting future problem.

We only study the four-dimensional $\Zb_2$ gauge theory, 
but our conclusions are expected to be the same for other theories with the same non-invertible symmetry.
For example, the four-dimensional Maxwell theory with the complex coupling $\tau=2i$ \cite{Choi:2021kmx}
and the $\mathcal{N}=4$ $SU(2)$ super Yang-Mills theory with $\tau=i$ have exactly the same symmetry \cite{Kaidi:2021xfk}.
Moreover, there are a lot of non-invertible symmetries in continuous quantum field theories in four dimensions.
It is an interesting future problem to study various boundary conditions and g-functions in these quantum field theories by using these non-invertible symmetries.
In this case, it should be a nice approach to find the fusion rules using the continuous field theory pictures as done in \cite{Choi:2021kmx,Kaidi:2021xfk,Choi:2022zal,Choi:2022jqy,Cordova:2022ieu}

It is also interesting to investigate non-topological defects and interfaces by using non-invertible symmetries.  There is also the g-theorem or g-conjecture related to defects and interfaces \cite{Nozaki:2012qd,Gaiotto:2014gha,Jensen:2015swa,Casini:2016fgb,Casini:2018nym,Kobayashi:2018lil,Wang:2021mdq,Cuomo:2021rkm,Shachar:2022fqk,Casini:2023kyj}.  In particular, some of them are known to have string theory duals.  Non-invertible symmetries are also investigated in the context of the AdS/CFT correspondence \cite{Apruzzi:2022rei,GarciaEtxebarria:2022vzq,Antinucci:2022vyk}.  Therefore, it will be interesting to study defects, interfaces and non-invertible symmetries in terms of branes in the string theory. 

\subsection*{Acknowledgment}
We would like to thank Dongmin Gang, Kentaro Hori, Justin Kaidi, Tatsuma Nishioka, Kantaro Ohmori, Soichiro Shimamori, Philip Boyle Smith, Yuji Tachikawa, Yuya Tanizaki, Hiroki Wada, Masahito Yamazaki, and Yunqin Zheng for helpful discussions.
SY would also like to thank the Yukawa Institute for Theoretical Physics at Kyoto University and Kavli Institute for the Physics and Mathematics of the Universe for hospitality during his stay. 
Discussions during the YITP workshop YITP-W-22-09 on ``Strings and Fields 2022'' were useful to complete this work.
SY was partially supported by Japan Society for the Promotion of Science (JSPS) Grant-in-Aid for Scientific Research Grant Number JP21K03574.

\appendix
\section{Commutation relations on \bbb}
\label{sec:commutation relations on bbb}
\begin{figure}[htbp]
  \begin{tabular}{ccc}
    \begin{minipage}[t]{0.45\hsize}
      \centering
      \includegraphics[width=5.5cm]{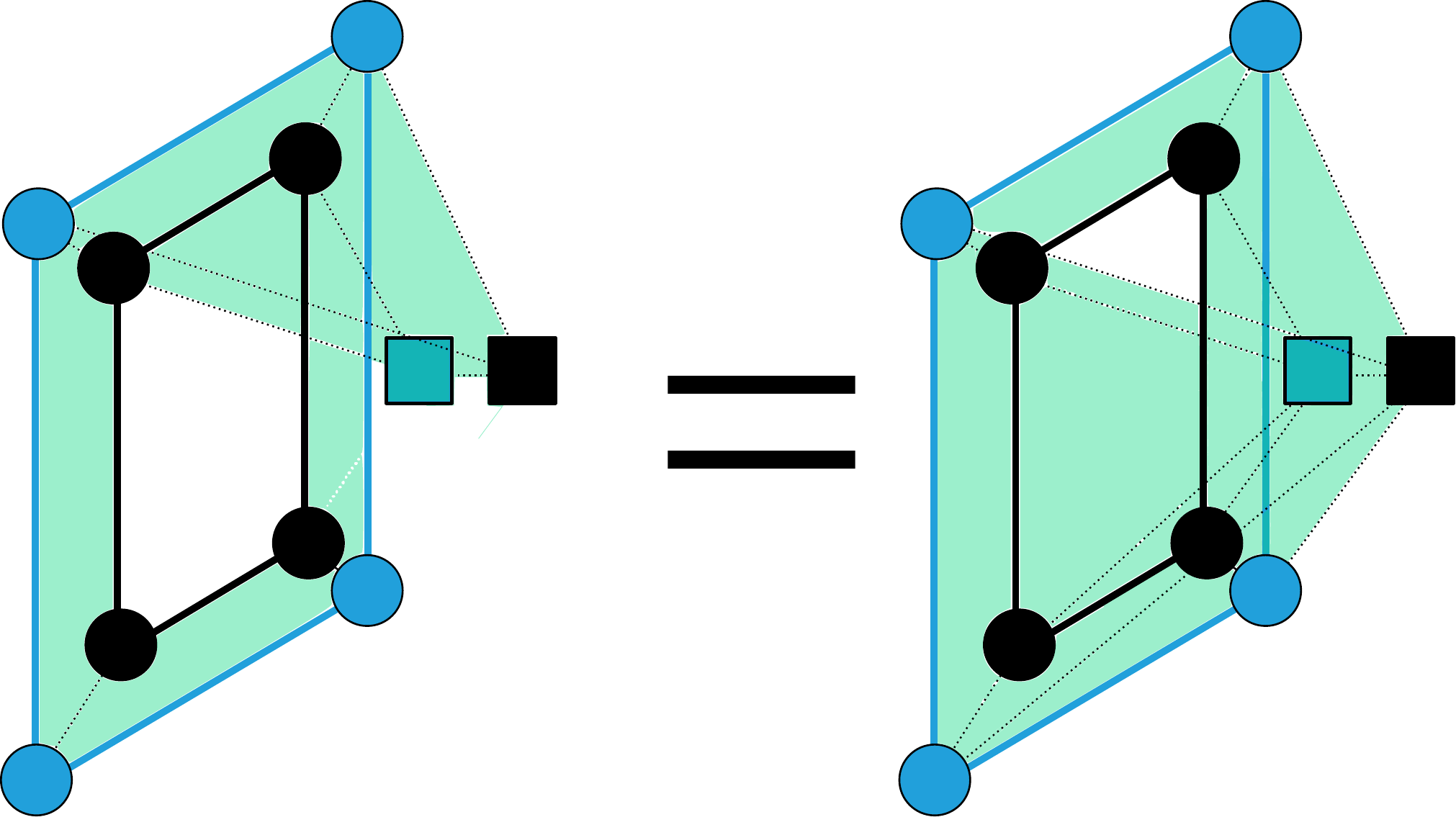}
      \subcaption{}
      \label{P1-3}
    \end{minipage} &
    \begin{minipage}[t]{0.45\hsize}
      \centering
      \includegraphics[width=5.5cm]{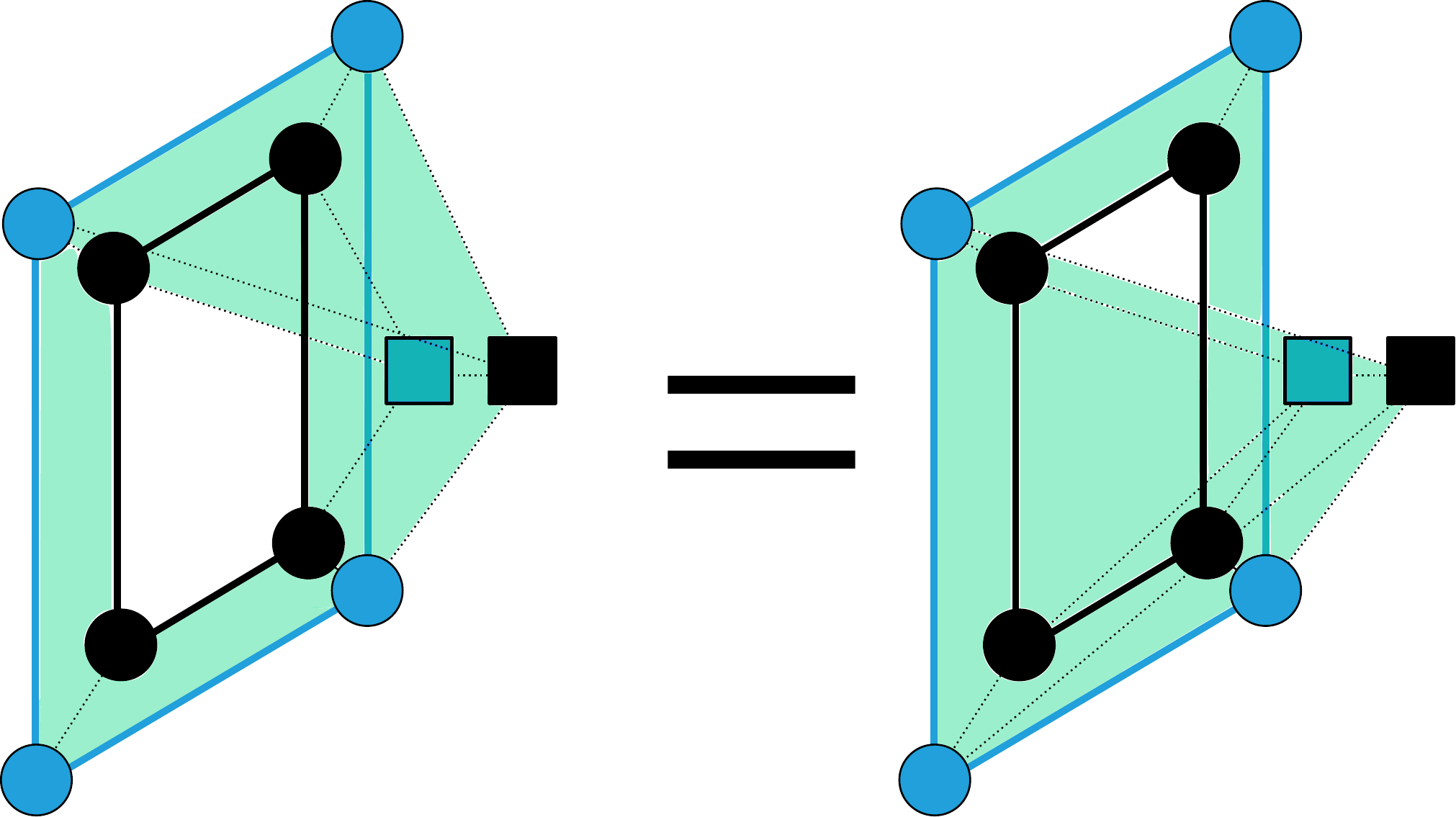}
      \subcaption{}
      \label{P2-2}
    \end{minipage} \\
    \begin{minipage}[t]{0.45\hsize}
      \centering
      \includegraphics[width=5.5cm]{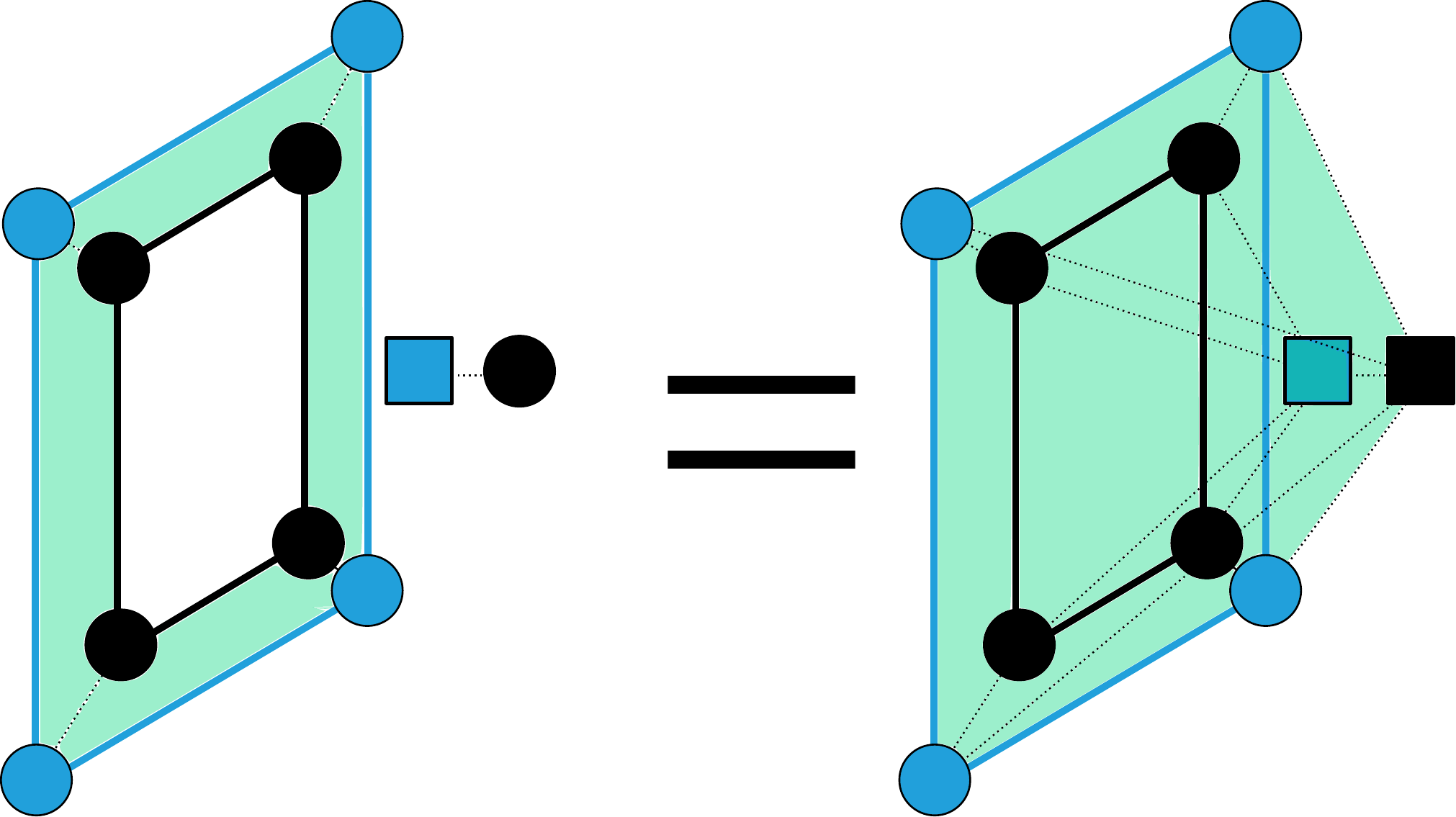}
      \subcaption{}
      \label{P0-4}
    \end{minipage} &
    \begin{minipage}[t]{0.45\hsize}
      \centering
      \includegraphics[width=5.5cm]{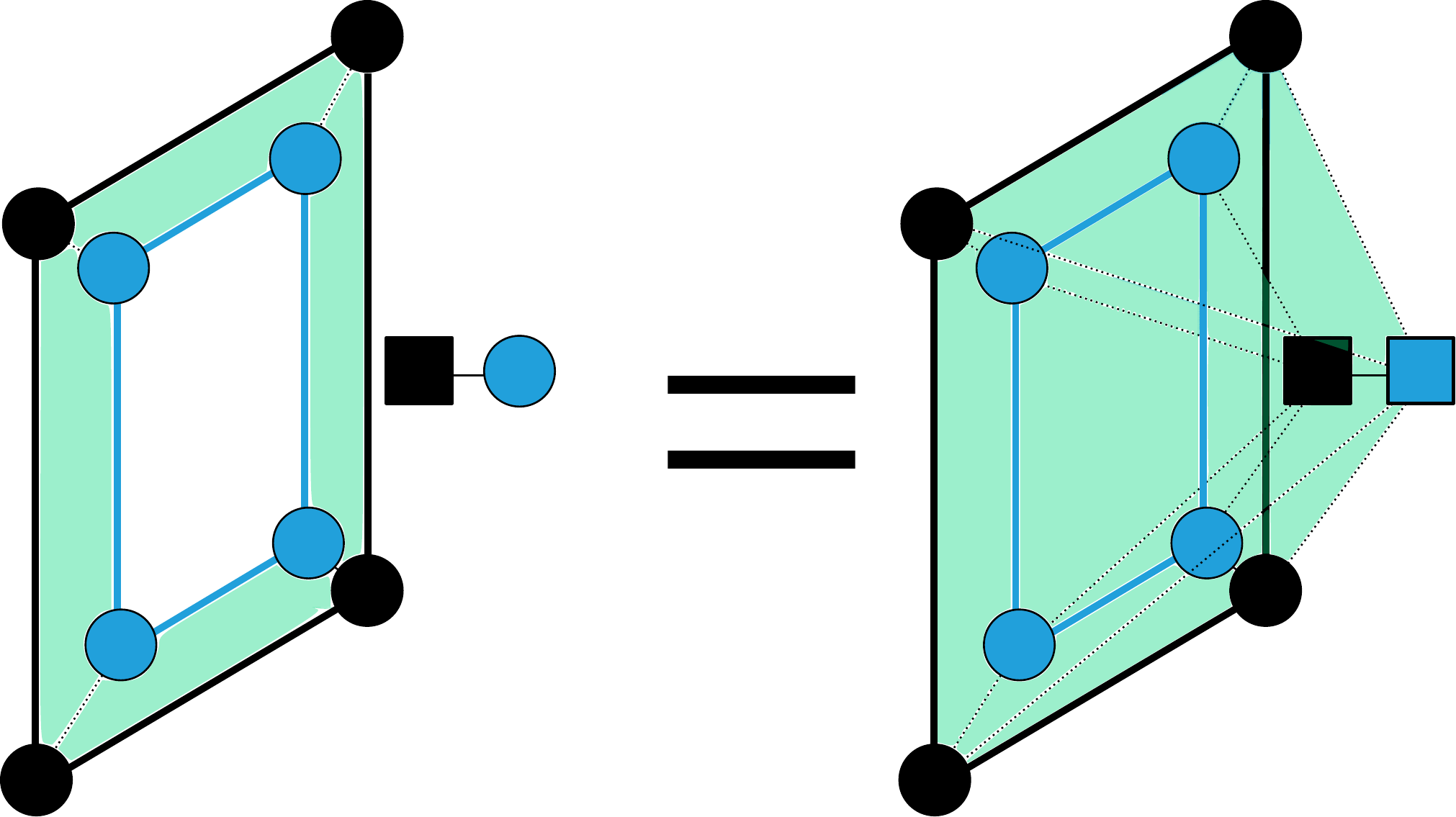}
      \subcaption{}
      \label{AP0-4}
    \end{minipage}
\end{tabular}
   \caption{Boundary defect commutation relations on a \bbb. Black plaquettes represent plaquettes on the boundary with D or \Dtb boundary condition and blue plaquettes represent plaquettes on the boundary with N boundary condition. Black square dots and blue square dots represent active links and inactive links in the bulk, respectively.  Black circular dots and blue circular dots represent active sites and inactive sites, respectively. Some of four tetrahedrons are filled by the KWW duality defect that are represented by green surfaces.  One square pyramid out of two is filled by the KWW defect thought it is omitted in the figure.}
   \label{fig:boundery defect commutation relations on a bbb}
\end{figure}

We study boundary defect commutation relations on a \bbb.
A \bbb contains four tetrahedrons and two square pyramids.
Let us consider a configuration of a KWW duality defect and a \bbb.
Suppose some tetrahedrons and one square pyramid on the surface of this \bbb are filled by the KWW duality defect.
We require that the partition function does not change even if the duality defect is deformed around the \bbb without changing the topology.
This condition is called a ``boundary defect commutation relation.''
There are four defect commutation relations of the KWW duality defect connecting D and N as shown in Figure~\ref{fig:boundery defect commutation relations on a bbb}.  There are also four defect commutation relations connecting \Dtb and N depicted by the same figures.

All of these boundary defect commutation relations are satisfied for arbitrary values of the weights of the elements on the boundary if we use the bulk weights obtained in \cite{Koide:2021zxj}.
For example, the boundary defect commutation relations of Figure~\ref{P0-4} connecting \Dtb and N reads
\begin{align}
  &\sum_{a_1,a_2,a_3,a_4}W_{\Dt}\deltamod{a_1+a_2+a_3+a_4}\pDt \bsDt^4\blDt^4s\notag\\
  &\qquad=\sum_{a_1,a_2,a_3,a_4,a_5}W_{\Dt}\deltamod{a_1+a_2+a_3+a_4}\pDt \bsDt^4\blDt^4s^2l^1D(a_5,a_1)D(a_5,a_2)D(a_5,a_3)D(a_5,a_4).
\end{align}
Here $a_1,\ a_2,\ a_3,\ a_4$ are link variables on the boundary, $a_5$ is a link variable in the bulk and $l=s=\frac{1}{\sqrt{2}}$ are weights in the bulk.
This equation is an identity with respect to $W_{\Dt},\pDt,\bsDt,\blDt$.

\bibliographystyle{utphys}
\bibliography{ref}
\end{document}